\documentclass[journal]{IEEEtran}

\ifCLASSINFOpdf
\else
   \usepackage[dvips]{graphicx}
\fi
\usepackage{url}

\usepackage{graphicx}
\usepackage{color}
\usepackage{diagbox}
\usepackage{multirow}
\usepackage{amsmath}
\usepackage{amssymb}
\usepackage{algorithm}
\usepackage{algpseudocode}
\usepackage{xcolor}
\usepackage{setspace}

\renewcommand{\vec}[1]{\boldsymbol{\mathrm{#1}}}

\begin{document}

\title{Optimizing a-DCF for Spoofing-Robust Speaker Verification}

\author{Oğuzhan Kurnaz, Jagabandhu Mishra, Tomi H. Kinnunen, and Cemal Hanilçi
\vspace{-0.86 cm}
\thanks{O. Kurnaz (Corresponding author), J. Mishra, and T. H. Kinnunen are from University of Eastern Finland, Joensuu, Finland. \\
O. Kurnaz and C. Hanilçi are from Bursa Technical University, Bursa, Turkey. \\
This study has been partially supported by the Academy of Finland (Decision No. 349605, project "SPEECHFAKES"). The authors wish to acknowledge CSC – IT Center for Science, Finland, for computational resources.}}

\markboth{Journal of \LaTeX\ Class Files, Vol. 14, No. 8, August 2015}
{Shell \MakeLowercase{\textit{et al.}}: Bare Demo of IEEEtran.cls for IEEE Journals}
\maketitle

\begin{abstract}
Automatic speaker verification (ASV) systems are vulnerable to spoofing attacks. We propose a spoofing-robust ASV system optimized directly for the recently introduced architecture-agnostic detection cost function (a-DCF), which allows targeting a desired trade-off between the contradicting aims of user convenience and robustness to spoofing. We combine a-DCF and binary cross-entropy (BCE) with a novel straightforward threshold optimization technique. Our results with an embedding fusion system on ASVspoof2019 data demonstrate relative improvement of $13\%$ over a system trained using BCE only (from minimum a-DCF of $0.1445$ to $0.1254$). Using an alternative non-linear score fusion approach provides relative improvement of $43\%$ (from minimum a-DCF of $0.0508$ to $0.0289$). 
\end{abstract}

\vspace{-0.4 cm}
\section{Introduction}

\emph{Automatic speaker verification} (ASV) seeks to verify a claimed identity based on speech evidence \cite{reynolds94_asriv}. Despite being a matured biometric technology, ASV 
is susceptible to \emph{spoofing} 
through replay \cite{kinnunen2017}, 
text-to-speech \cite{todisco2019asvspoof} 
and other attacks. To counter these risks, various \emph{countermeasures} (CMs) or \emph{presentation attack detectors} (PADs) \cite{ISO}, have been proposed 
for detecting spoofing.
Typically CMs are 
developed in isolation from ASV, and the two subsystems 
are combined to enhance security. 
The common strategies include tandem (cascade) combination \cite{kinnunen2020tandem} and fusion of detection scores \cite{Todisco2018} or embeddings \cite{Shim2022,Liu2024}. Besides systems constructed from combination of speaker and spoof detection subsystems, 
monolithic end-to-end solutions 
\cite{kang22_interspeech, Teng2022}, \cite{mun23_interspeech} have also been considered. 


Whether realized as a combination of the ASV and CM subsystems, 
or an end-to-end model, a key consideration is how to evaluate and optimize the complete 
system, taking into account the 
trade-off between misses (false rejections) and false alarms (false acceptances)---proxies for user (in)convenience and security, respectively. The 
so-called \emph{tandem detection cost function} (t-DCF) 
\cite{kinnunen2020tandem} can be used to evaluate 
CM-ASV cascades 
under a Bayes-risk framework. Nonetheless, t-DCF requires two sets of detection scores (and two thresholds) corresponding to the two subsystems, making it inapplicable for 
evaluation of any other type of architecture.  
To address this limitation,
\cite{shim2024adcf} proposed recently an 
\emph{architecture-agnostic detection cost function} (a-DCF) applicable to a wider variety of architectures. 
Unlike t-DCF, a-DCF requires only one set of detection scores and one 
detection threshold. 
This new metric was the primary metric in one of the two main tracks in the ASVspoof 5 challenge \cite{Wang2024_ASVspoof5}.


\begin{figure}[!t]
\centering

\includegraphics[scale=0.3]{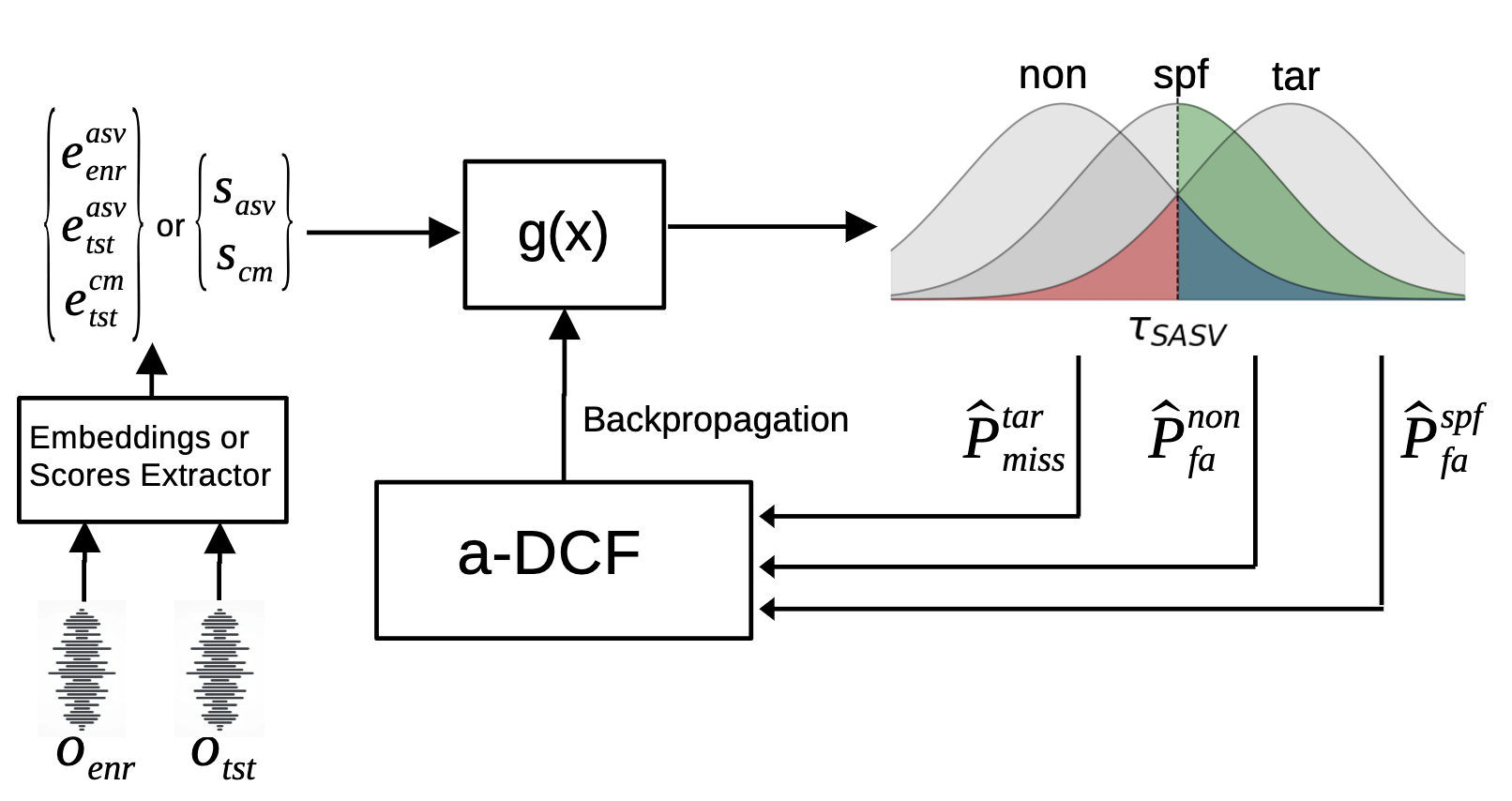}
\vspace{-0.4cm}
\caption{The proposed a-DCF optimization for spoofing-robust speaker verification uses 
embeddings or scores from enrollment ($\mathcal{O}_{\text{enr}}$) and test ($\mathcal{O}_{\text{tst}}$) inputs 
to give scores $g(x)$ 
for target, non-target, and spoof classes. Error rates $\hat{P}_\mathrm{miss}^\mathrm{tar}$, $\hat{P}_\mathrm{fa}^\mathrm{non}$, and $\hat{P}_\mathrm{fa}^\mathrm{spf}$, computed using threshold $\tau_{\text{sasv}}$, form the soft a-DCF loss for backpropagation to optimize class-specific detection accuracy.}
\vspace{-0.4cm}
\label{fig:adcf_model}
\end{figure}

\textbf{For the first time, we address spoofing-robust ASV 
directly optimized for the a-DCF metric} (Fig.~\ref{fig:adcf_model}). As a 
metric based on 
hard error counting, 
a-DCF 
is non-differentiable and hence cannot be directly optimized using gradient descent schemes. Luckily, hard error counts can be ``softened'' (made differentiable). 
While direct optimization of the t-DCF was addressed in \cite{kanervisto2021optimizing} through reinforcement learning, this approach is limited to the tandem architectures. By integrating a differentiable version of the a-DCF into model training, we aim to enhance the performance and robustness of speaker verification systems against spoofing attacks, ultimately contributing to the advancement of secure and reliable biometric authentication technologies. Besides integrating the a-DCF metric into model training, the threshold 
($\tau_{\mathrm{sasv}}$) is jointly optimized with the 
model parameters 
in the proposed method.

\vspace{-0.30cm}
\section{Detection Cost Functions}


\emph{Detection cost functions} \cite{Doddington2000-NIST} are evaluation metrics used to assess the performance of binary classifiers, such as speaker verification and anti-spoofing systems. 
A binary classifier can be defined as a function $ g: \mathcal{X} \to \mathbb{R}$ 
that assigns a detection score $g(x)$ for an input $ x \in \mathcal{X} $, 
converted subsequently into predicted 
label $y \in \{0,1\}$ by comparing it with a 
threshold $\tau$, i.e. $y = \mathrm{I}( g(x) > \tau )$. Here, $\mathrm{I}(\cdot)$ is the indicator function which returns $1$ if the condition is true (null hypothesis accepted) and $0$ otherwise. In 
ASV, $x=(\mathcal{O}_\text{enr},\mathcal{O}_\text{tst})$ consists of a pair of enrollment $(\mathcal{O}_\text{enr})$ and test $(\mathcal{O}_\text{tst})$ utterances. The null hypothesis is that the same speaker produced both utterances. To develop a spoofing-aware speaker verification (SASV) system, we represent $x$ either as a triplet of multivariate embeddings $(\vec{e}_\text{enr}^\text{asv},\vec{e}_\text{tst}^\text{asv},\vec{e}_\text{tst}^\text{cm})$, or as a pair of scalars 
$(s_\text{asv},s_\text{cm})$. In the former case, the two 
\emph{speaker embeddings }$(\vec{e}_\text{enr}^\text{asv},\vec{e}_\text{tst}^\text{asv})$ extracted using an ASV model are augmented by a 
\emph{spoof embedding} $\vec{e}_\text{tst}^\text{cm}$ extracted using a CM model. The latter case, 
consists of a pair of ASV and CM scores 
We demonstrate our back-end optimization for both types of input representations. 
\vspace{-0.35cm}
\subsection{DCF and a-DCF}

\noindent \textbf{Detection Cost Function (DCF}) \cite{Doddington2000-NIST} is used for performance assessment of conventional ASV systems. It 
considers consequences of both false alarms (FAs) and misses through a cost model. 
The DCF is formally defined as,
\begin{equation}
\footnotesize
\label{dcf}
\mathrm{DCF(\tau_{asv})} = C_{\mathrm{miss}}^{\mathrm{tar}} \cdot \pi_{\mathrm{tar}} \cdot P_{\mathrm{miss}}^{\mathrm{tar}}(\tau_{\mathrm{asv}}) + C_{\mathrm{fa}}^{\mathrm{non}} \cdot  \pi_{\mathrm{non}} \cdot P_{\mathrm{fa}}^{\mathrm{non}}(\tau_{\mathrm{asv}}) , 
\end{equation}

\noindent where \( C_{\mathrm{miss}}^{\mathrm{tar}} \) and \( C_{\mathrm{fa}}^{\mathrm{non}} \) are the costs of a miss and FA, respectively, \( P_{\mathrm{miss}}^{\mathrm{tar}} \) and, \( P_{\mathrm{fa}}^{\mathrm{non}} \) are the miss and FA rates, \( \pi_{\mathrm{tar}} \) and \( \pi_{\mathrm{non}} = 1 - \pi_{\mathrm{tar}} \) represent target and nontarget prior probabilities, and $\tau_\mathrm{asv}$ is the detection threshold. The DCF in Eq. \eqref{dcf} allows tuning of cost parameters to specific application needs.


\textbf{Architecture-agnostic detection cost function (a-DCF)} \cite{shim2024adcf} is used for assessing spoofing-robust ASV systems. It 
extends 
the DCF by including a third class (spoofing attack):
\begin{equation}
\footnotesize
\label{adcf_old}
\begin{aligned}
\mathrm{a\!-\!DCF}(\tau_\mathrm{sasv}) = & \; C_{\mathrm{miss}}^{\mathrm{tar}} \cdot \pi_{\mathrm{tar}} \cdot P_{\mathrm{miss}}^{\mathrm{tar}}(\tau_\mathrm{sasv}) \\
& + C_{\mathrm{fa}}^{\mathrm{non}} \cdot \pi_{\mathrm{non}} \cdot P_{\mathrm{fa}}^{\mathrm{non}}(\tau_\mathrm{sasv}) \\
& + C_{\mathrm{fa}}^{\mathrm{spf}} \cdot \pi_{\mathrm{spf}} \cdot P_{\mathrm{fa}}^{\mathrm{spf}}(\tau_\mathrm{sasv}),
\end{aligned}
\end{equation}

\noindent where \( P_{\mathrm{fa}}^{\mathrm{spf}} \), 
\( \pi_{\mathrm{spf}} \), 
\( C_{\mathrm{fa}}^{\mathrm{spf}} \) and $\tau_\mathrm{sasv}$ 
are the FA rate of spoofs, prior of spoofing, cost of spoof FA, 
and the threshold, respectively. 
The three error rates in \eqref{adcf_old} 
are computed as follows:

\begin{equation}
\footnotesize
\begin{aligned}
P_{\mathrm{miss}}^{\mathrm{tar}}(\tau_\mathrm{sasv}) &= \frac{1}{N_\mathrm{tar}} \sum_{x \in \mathrm{tar}} \mathrm{I}(g(x) \leq \tau_\mathrm{sasv}) \\
P_{\mathrm{fa}}^{\mathrm{non}}(\tau_\mathrm{sasv}) &= \frac{1}{N_\mathrm{non}} \sum_{x \in \mathrm{non}} \mathrm{I}(g(x) > \tau_\mathrm{sasv}) \\
P_{\mathrm{fa}}^{\mathrm{spf}}(\tau_\text{sasv}) &= \frac{1}{N_\mathrm{spf}} \sum_{x \in \mathrm{spf}} \mathrm{I}(g(x) > \tau_\mathrm{sasv}),
\end{aligned}
\label{eq:all_p}
\end{equation}
where $\mathrm{tar}$, $\mathrm{non}$ and $\mathrm{spf}$ denote the sets of target, non-target, and spoof trials, respectively, with their counts denoted by $N_\bullet$.

\vspace{-0.2cm}

\subsection{Differentiable a-DCF Metric}
The indicator function 
in Eq. \eqref{eq:all_p} is non-differentiable. Therefore, 
the a-DCF metric can not be directly used as 
loss function in gradient search methods. 
In this work, we propose a differentiable (soft) version of a-DCF suitable for this purpose. 
Similar approaches have been successfully used to optimize DCF for text-dependent speaker verification \cite{mingote19_interspeech} and t-DCF for tandem architectures \cite{kanervisto2021optimizing}. Compared with t-DCF, applicable only to the unnecessarily restrictive class of tandem systems, a-DCF is applicable to arbitrary SASV systems. Refer to \cite{shim2024adcf} for further elaboration.

The non-differentiable components in a-DCF metric are the miss rate ($P_{\text{miss}}$) and the two false alarm rates ($P_{\text{fa}}^{\text{non}}$, $P_{\text{fa}}^{\text{spf}}$). A common approach for `softening' hard error counts \cite{mingote19_interspeech,kanervisto2021optimizing} is to replace the indicator function by a continuous, differentiable function. We adopt the following approximations:


\begin{equation}
\footnotesize
\begin{aligned}
\hat{P}_{\mathrm{miss}}^{\mathrm{tar}}(\tau_\mathrm{sasv}) &= \frac{1}{N_\mathrm{tar}} \sum_{x \in \mathrm{tar}} \sigma \left( \tau_\mathrm{sasv} - g(x)\right) \\
\hat{P}_{\mathrm{fa}}^{\mathrm{non}}(\tau_\mathrm{sasv}) &= \frac{1}{N_\mathrm{non}} \sum_{x \in \mathrm{non}} \sigma \left( g(x) - \tau_\mathrm{sasv}\right) \\
\hat{P}_{\mathrm{fa}}^{\mathrm{spf}}(\tau_\mathrm{sasv}) &= \frac{1}{N_\mathrm{spf}} \sum_{x \in \mathrm{spf}} \sigma \left( g(x) - \tau_\mathrm{sasv} \right),
\end{aligned}
\label{eq:all_p_hat}
\end{equation}


\noindent where $\sigma(z) = \frac{1}{1 + e^{-z}}$ is the sigmoid function. Essentially, the error rates are approximated based on their distance from the threshold. 
The soft a-DCF loss is obtained by replacing the error rate terms in Eq. \eqref{eq:all_p} with the expressions from Eq. \eqref{eq:all_p_hat}, applied within the original a-DCF formula in Eq. \eqref{adcf_old}.



\vspace{-0.3cm}
\subsection{Optimization of soft a-DCF}
\vspace{-0.1cm}

\begin{algorithm}
\small
\caption{Optimization Algorithm}
\label{algrthm}
\begin{algorithmic}[1]
\State \textbf{Input:} Train data $\mathcal{D}_{\mathrm{trn}}$, Development data $\mathcal{D}_{\mathrm{dev}}$, Batch Size $B$, Number of Epochs $N$
\State \textbf{Output:} Best threshold $\tau_{\mathrm{sasv}}$, Model Parameters $\Theta_{*}$
\State \textbf{Initialize:} $\mathrm{min\_a\!-\!DCF} \gets \infty$, $\tau \gets 0.5$
\For{i $= 1$ to $N$}
    \State \textcolor{blue}{// optimize DNN weights}
    \For{$X \gets get\_minibatch(\mathcal{D}_{\mathrm{trn}}, B)$}
        \State $\mathcal{J}(\Theta_{i}, \tau_{i}) \gets \left( \mathcal{L}_{\mathrm{a-DCF}}^{\mathrm{soft}}(X; \Theta_{i}, \tau_{i}) + \mathcal{L}_{\mathrm{BCE}}(X; \Theta_{i}) \right) / 2$
        \State $\Theta_{i+1} \gets update\_network\_weights(\Theta_i, \mathcal{J}(\Theta_i, \tau_i))$
    \EndFor
    \State $\Theta \gets \Theta_{i+1}$
    \State \textcolor{blue}{// find optimal threshold using grid search}
    \State $\tau_{i+1} \gets \operatorname*{argmin}\limits_\tau \{\mathcal{L}_{\mathrm{a-DCF}}^{\mathrm{soft}}(\Theta, \tau)\}$ in $\mathcal{D}_{\mathrm{trn}}$
    \State \textcolor{blue}{// keep track of best model using dev data}
    \If{$\mathcal{L}_{\mathrm{a-DCF}}^{\mathrm{soft}}(\tau_{i+1}) < \mathrm{min\_a\!-\!DCF}$ in $\mathcal{D}_{\mathrm{dev}}$}
        \State $\mathrm{min\_a\!-\!DCF} \gets \mathcal{L}_{\mathrm{a-DCF}}^{\mathrm{soft}}(\tau_{i+1})$
        \State $\tau_{\mathrm{sasv}} \gets \tau_{i+1}$
        \State $\Theta_{*} \gets \Theta$
    \EndIf
\EndFor
\State \textbf{return} $\tau_{\mathrm{sasv}}$, $\Theta_{*}$
\end{algorithmic}
\end{algorithm}

The algorithm \ref{algrthm} presents the optimization loop of a 
generic model through a combination of soft a-DCF ($\mathcal{L}_{\mathrm{a-DCF}}^{\mathrm{soft}}(X; \Theta_{i}, \tau_{i})$) and BCE  ($\mathcal{L}_{\mathrm{BCE}}(X; \Theta_{i})$) losses. The former is defined as, 
\vspace{-0.1cm}
\begin{equation}
\footnotesize
  \mathcal{L}_\mathrm{BCE} = -\frac{1}{N} \sum_{j=1}^{N} [y_j \log(g(x_j)) + (1 - y_j) \log(1 - g(x_j))],
\end{equation}

\noindent where $N$ is the number of training samples and $g(x_j)$ is the model output (score). The algorithm initializes a minimum a-DCF to be infinity and a fixed threshold $\tau$. Then, for each epoch, it optimizes the model weights by calculating the combined loss $\mathcal{J}(\Theta_{i}, \tau_{i})$ from soft a-DCF and BCE losses and updating the weights of the neural network. This combined loss is defined as Line 7 in Algorithm~\ref{algrthm}.

The algorithm then conducts a grid search (Line 12 in Algorithm \ref{algrthm}) for a value of $\tau$ that gives the optimal threshold according to the a-DCF loss on the training data. The threshold value serves as the independent variable in Eq. \eqref{eq:all_p_hat} of the a-DCF loss. Using the a-DCF loss (or its combination with BCE loss) as the objective function during model training involves minimizing its value, which depends on the threshold value. This relationship necessitates finding the optimal threshold value that minimizes the a-DCF loss at each epoch which is achieved through grid search. This algorithm tracks the best model parameters and threshold as scored on development data, thus updating the lowest a-DCF found so far and its corresponding parameters. At the end, it returns the best threshold $\tau_{\mathrm{sasv}}$



To elaborate on the combined loss in Line 7, assume that the empirical proportion of the training data from each class (target, nontarget, spoof) are equal ($\sim \frac{1}{3}$) in each minibatch. 
Expanding the combined loss $\mathcal{J}(\Theta_{i}, \tau_{i})$, 
after some algebraic manipulation, can be shown to give
\begin{equation}
\footnotesize
\begin{aligned}\label{eq:combined-loss}
    \mathcal{J} \propto & \frac{1}{N_{\text{tar}}} \sum_{x \in\text{tar}} \left[ A \sigma \left( \tau_{\text{sasv}} - g(x) \right) - \log g(x) \right] \\
    & + \frac{1}{N_{\text{non}}} \sum_{x \in\text{non}} \left[ B \sigma \left( g(x) - \tau_{\text{sasv}} \right) - \log (1 - g(x)) \right] \\
    & + \frac{1}{N_{\text{spf}}} \sum_{x \in\text{spf}} \left[ C \sigma \left( g(x) - \tau_{\text{sasv}} \right) - \log (1 - g(x)) \right],
\end{aligned}
\end{equation}
\noindent where $A$, $B$ and $C$ are 
constants implied by the a-DCF parameters\footnote{
In specific, $A := 3C_{\text{tar}}^{\text{miss}} \pi_{\text{tar}}$, $B := 3C_{\text{non}}^{\text{fa}} \pi_{\text{non}}$, and $C := 3C_{\text{spf}}^{\text{fa}} \pi_{\text{spf}}$}. and where $\propto$ signifies omission of unimportant scaling constants. The combined loss is 
balanced average of the three class-conditional 
loss terms, with the per-instance losses enclosed within the brackets. The relative contribution of the two types of terms depends on the a-DCF parameters. While the sigmoid (from a-DCF) and the log (from BCE) terms are tied through the score $g(x)$, $\tau_\text{sasv}$ is present only in the a-DCF terms---but tied across the class-conditional losses.

As we will demonstrate below, the interplay of the two losses is important in 
optimization. While BCE focuses on class separation (discrimination), 
a-DCF takes into account the costs of \emph{actual decisions} by including the decision threshold. 
Together, BCE and a-DCF form a complementary loss function that not only establishes a robust classification boundary but also adapts the model to prioritize specific classes based on their operational relevance and associated costs.


\vspace{-0.25cm}

\section{Experimental Setup, Results and Discussion}
\subsection{Dataset}



The SASV experiments are conducted on the ASVspoof2019 logical access (LA) dataset \cite{todisco2019asvspoof}, which consists of training, testing, and evaluation subsets, with disjoint speaker identity. This dataset includes both bonafide and spoofed utterances, with spoofed signals generated using various voice conversion (VC) and text-to-speech (TTS) algorithms. In these experiments, embeddings extracted using the pretrained ECAPA-TDNN and AASIST models provided by the organizers for the ASV and CM tasks \cite{Shim2022} were utilized. A detailed dataset description is available at \cite{todisco2019asvspoof}.



\vspace{-0.3cm}
\subsection{Experimental Setup}

To demonstrate our proposed optimization scheme, we consider the DNN-based embedding fusion strategy 'Baseline2' described in \cite{Shim2022}, 
used as a baseline in the SASV2022 challenge. Thereafter, many other studies follow similar approach to perform SASV \cite{Zhang2022, zhang2022backend}. 
The two speaker embeddings 
$(\textbf{e}_\text{enr}^\text{asv},\textbf{e}_\text{tst}^\text{asv})$ 
are extracted through ECAPA-TDNN \cite{Desplanques_2020} and the spoof embedding ($\textbf{e}_\text{tst}^\text{cm}$) 
through AASIST \cite{jung2022aasist}. The three concatenated embeddings are fed to a DNN model that consists of three fully connected hidden layers of $256$, $128$, and $64$ neurons with leaky ReLU activation functions. Unlike the baseline’s two-neuron softmax output, our model's final layer has a single neuron with a sigmoid activation, as a-DCF optimization requires only one output score.


We consider four back-end models with identical architectures, optimized differently. They are: 
\begin{itemize}
    \item \textbf{S1}: corresponds to the Baseline2 system provided by SASV challenge organizers (optimized with CE);
    \item \textbf{S2}: optimized using soft a-DCF, with $\tau_\mathrm{sasv} = 0.5$; 
    \item \textbf{S3}: optimized using soft a-DCF + BCE, with $\tau_\mathrm{sasv} = 0.5$;
    \item \textbf{S4}: optimized using soft a-DCF + BCE, including threshold optimization, as detailed in Algorithm \ref{algrthm}.
\end{itemize}
In addition to these embedding fusion methods, 
we 
proposed a-DCF optimization 
of a trainable 
nonlinear score fusion method \cite{wang24l_interspeech} (detailed below). 


The cost values \( C_{\mathrm{miss}}^{\mathrm{tar}} \), \( C_{\mathrm{fa}}^{\mathrm{non}} \), and \( C_{\mathrm{fa}}^{\mathrm{spf}} \) in the soft a-DCF calculation are set as $1$, $10$, and $20$, respectively, while the prior values \( \pi_{\mathrm{tar}} \), \( \pi_{\mathrm{non}} \), and \( \pi_{\mathrm{spf}} \) are set as $0.9$, $0.05$, and $0.05$. The higher relative cost value for spoof false alarms (\( C_{\mathrm{fa}}^{\mathrm{spf}} = 20 \)) is chosen based on the importance of spoofing robustness in security-sensitive domains, such as banking, where preventing unauthorized access is crucial. This reflects the emphasis to penalize false acceptance of spoof attempts to ensure the integrity and security of such systems. Additionally, the high target prior (\( \pi_{\mathrm{tar}} = 0.9 \)) 
aligns with an assumption that most interactions are legitimate. 
The a-DCF metric in \eqref{adcf_old} is used for performance evaluation. Additionally, 
\emph{equal error rate} (EER) is reported in selected cases. 
We use SV-EER (speaker verification EER) to assess the system's ability to differentiate between target and non-target samples, and SPF-EER (spoofing EER) 
to assess ability to differentiate between bonafide targets and spoofing attacks.

\vspace{-0.2cm}

\begin{table}[h!]
\centering
\vspace{-0.15cm}
\caption{Baseline system min \emph{a-DCF} values in the development set}
\fontsize{5.5 pt}{8 pt}\selectfont
\begin{tabular}{|c|c|c|c|c|c|c|c|c|c|}
\hline
Batch size & 24 & 48 & 96 & 192 & 384 & 768 & 1024 & 2048 \\ \hline
min a-DCF & 0.1437 & 0.1308 & 0.1268 & 0.1346 & 0.1297 & 0.1251 & \textbf{0.1234} & 0.1243 \\ \hline
\end{tabular}
\vspace{-0.6cm}
\label{tab:baseline_dev}
\end{table}


\subsection{Baseline (\textbf{S1})}

We firstly trained a baseline (\textbf{S1}) with cross entropy loss for various batch sizes, resulting to minimum a-DCF values shown in Table~\ref{tab:baseline_dev}. The results indicate that the best performance was achieved with a batch size of $1024$. We fix this value for the remainder of the experiments.




\vspace{-0.25cm}

\subsection{Proposed a-DCF metric optimization (\textbf{S2}, \textbf{S3} and \textbf{S4})}
We next consider models \textbf{S2} and \textbf{S3}, beginning with the hard-coded detection threshold of $\tau_\mathrm{sasv} = 0.5$ in the soft a-DCF loss equation. Model selection is based on the lowest a-DCF value at a threshold of $0.5$ in the development set. The soft a-DCF only system (\textbf{S2}) attains a minimum a-DCF of $0.1355$, with an SV-EER of $9.76\%$ and an SPF-EER of $0.14$ in the development set, while in the evaluation set, it reaches a minimum a-DCF of $0.2352$, an SV-EER of $17.28\%$ and an SPF-EER of $1.02\%$. In contrast, the combination of the soft a-DCF and BCE (\textbf{S3}) achieves a minimum a-DCF of $0.1182$, with an SV-EER of $8.29\%$ and an SPF-EER of $0.10\%$ in the development set, and in the evaluation set, it attains a minimum a-DCF of $0.1398$, an SV-EER of $9.59\%$ and an SPF-EER of $0.63\%$, indicating improved accuracy. 



\begin{table}[h!]
\footnotesize
\centering
\caption{Comparison of \textbf{S1}, \textbf{S2}, \textbf{S3}, and \textbf{S4} in development and evaluation set. The results are given using 1024 as batch size.}
\begin{tabular}{|c|c|c|c|c|c|c|}
\hline
\multirow{2}{*}{System} & \multicolumn{2}{c|}{a-DCF} & \multicolumn{2}{c|}{SV-EER} & \multicolumn{2}{c|}{SPF-EER} \\ \cline{2-7} 
                    & Dev & Eval & Dev & Eval & Dev & Eval \\ \hline
\textbf{Baseline (S1)} & 0.1234 & 0.1445 & 8.36 & 9.86 & \textbf{0.07} & 0.63 \\ \hline
\textbf{S2} & 0.1355 & 0.2352 & 9.76 & 17.28 & 0.14 & 1.02 \\ \hline
\textbf{S3} & 0.1182 & 0.1398 & 8.29 & 9.59 & 0.10 & 0.63 \\ \hline
\textbf{S4} & \textbf{0.1109} & \textbf{0.1254} & \textbf{7.75} & \textbf{8.44} & 0.08 & \textbf{0.61} \\ \hline
\end{tabular}
\label{tab:all_systems_dev_eval}
\end{table}


The results of \textbf{S2} and \textbf{S3} were based on the hard-coded arbitrary detection threshold of $\tau_\mathrm{sasv} = 0.5$. We now also optimize $\tau_\mathrm{sasv}$ using the complete procedure (Algorithm~\ref{algrthm}). Joint optimization of the DNN parameters and the threshold leads to noticeable improvements, as expected. Specifically, \textbf{S4} achieved a minimum a-DCF of $0.1109$, with an SV-EER of $7.75\%$ and an SPF-EER of $0.08$ on the development set, and in the evaluation set, it reached a minimum a-DCF of $0.1254$, an SV-EER of $8.44\%$ and an SPF-EER of $0.61\%$, outperforming \textbf{S1}, \textbf{S2}, and \textbf{S3}, as Table~\ref{tab:all_systems_dev_eval} indicates.


\begin{figure}[!t]
\centering
\vspace{-0.43cm}
\includegraphics[scale=0.47]{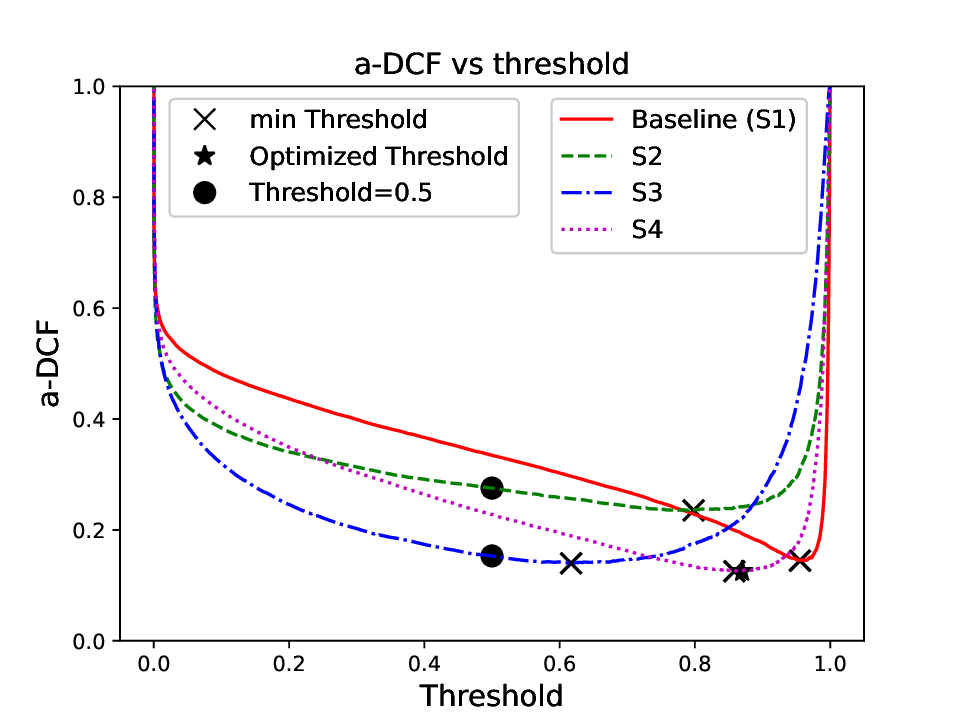}
\vspace{-0.2cm}
\caption{a-DCF values:\emph{Baseline ({\bf S1})}, \emph{\bf S2}, \emph{\bf S3}, and \emph{\bf S4} across different thresholds.} 
\vspace{-0.52cm}
\label{fig:adcf_plot}
\end{figure}

Fig.~\ref{fig:adcf_plot} presents the a-DCF curves for four systems (\textbf{S1}, \textbf{S2}, \textbf{S3}, and \textbf{S4}) across varying thresholds. For \textbf{S2} and \textbf{S3}, both the a-DCF values at the fixed threshold of $0.5$ and the minimum points are indicated. The \textbf{S4} system, which incorporates threshold optimization, has its optimized a-DCF value marked with a star. As seen in the graph, \textbf{S4} yields the lowest a-DCF value.


\vspace{-0.4cm}
\subsection{Impact of a-DCF Parameter}



\begin{figure}[h]
\vspace{-0.45cm}
\centering
\includegraphics[scale=0.25]{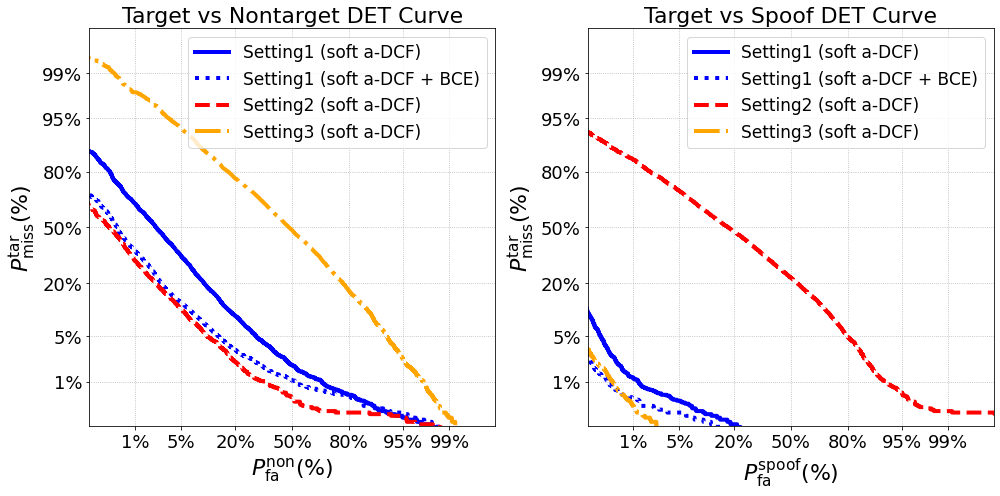}
\vspace{-0.7cm}
\caption{DET curves for three a-DCF parameter settings. SV-EER values are $\textbf{13.97\%}$ for \emph{Setting1} (solid blue), $\textbf{8.44\%}$ for \emph{Setting1} (dotted blue), $\textbf{7.52\%}$ for \emph{Setting2}, and $\textbf{49.13\%}$ for \emph{Setting3}.}
\vspace{-0.2cm}
\label{fig:ops_sys}
\end{figure}

Our final analysis addresses the impact of the a-DCF parameters. To this end, we introduced three different sets of a-DCF parameters as:

\begin{itemize}
    \item \textbf{Setting1 (soft a-DCF)} contains only a-DCF loss. In this case, the cost values \( C_{\mathrm{miss}}^{\mathrm{tar}} \), \( C_{\mathrm{fa}}^{\mathrm{non}} \), and \( C_{\mathrm{fa}}^{\mathrm{spf}} \) are set to $1$, $10$, and $20$, respectively. The prior values \( \pi_{\mathrm{tar}} \), \( \pi_{\mathrm{non}} \), and \( \pi_{\mathrm{spf}} \) are set to $0.9$, $0.05$, and $0.05$. 
    \item \textbf{Setting1 (soft a-DCF + BCE)} contains both a-DCF and BCE losses. The cost and the prior values are the same as in the previous case. 
    \item \textbf{Setting2 (soft a-DCF)} contains only a-DCF loss. In this case, the cost values \( C_{\mathrm{miss}}^{\mathrm{tar}} \), \( C_{\mathrm{fa}}^{\mathrm{non}} \), and \( C_{\mathrm{fa}}^{\mathrm{spf}} \) are all set to $1$. The prior values \( \pi_{\mathrm{tar}} \), \( \pi_{\mathrm{non}} \), and \( \pi_{\mathrm{spf}} \) are set to $0.5$, $0.5$, and $0$, respectively.
    \item \textbf{Setting3 (soft a-DCF)} contains only a-DCF loss. The same cost values are used as in the previous setting, and $0.5$, $0$ and $0.5$ priors are used for \( \pi_{\mathrm{tar}} \), \( \pi_{\mathrm{non}} \), and \( \pi_{\mathrm{spf}} \), respectively. 
\end{itemize}


The threshold optimization method was included for all these cases. Fig. \ref{fig:ops_sys} shows target vs. nontarget and target vs. spoof DET profiles for three different sets of a-DCF parameters. 
The choice of a-DCF parameters significantly impacts both DET profiles. Comparing \emph{setting1}'s soft a-DCF alone with the soft a-DCF combined with BCE, the former generally has a lower miss rate. For \emph{setting2} and \emph{setting3}, changes in the priors cause notable differences in the DET curves. For example, setting $\pi_{\mathrm{spf}}$ to $0$ (\emph{setting2}), assuming no spoof non-targets, shows the best performance for target vs. nontarget trials across all operating points. However, when the non-target prior is set to $0$ (\emph{setting3}), the performance for target vs. nontarget trials is found to be inferior, as expected. These trends are reversed for target vs. spoof trials.

\subsection{a-DCF Loss in Score Fusion}



Finally, we demonstrated SASV performance using score fusion (SF) optimized with soft a-DCF. As reported in~\cite{wang24l_interspeech}, nonlinear calibrated score fusion (\textbf{NL-Cal-SF}) outperforms 
both linear sum (\textbf{L-SF}) and linear calibrated 
sum (\textbf{L-Cal-SF}). 
 We therefore applied soft a-DCF optimization with nonlinear score fusion (\textbf{aDCF-NL-Cal-SF}) of the form $-\log[\rho \cdot \text{exp}(-a s_\text{asv} + b)+(1-\rho) \cdot \text{exp}(-c s_\text{cm} + d)]$, where $\rho=0.5$ and $a$, $b$, $c$, $d$ are trainable parameters. Results and trainable parameter counts 
in Table~\ref{tab:score_fusion_tab}, 
indicate that 
\textbf{aDCF-NL-Cal-SF} improves 
over \textbf{NL-Cal-SF}. 
Furthermore, 
SF achieves better 
performance compared with embedding fusion, despite 
significantly fewer parameters ($4$ vs $180736$).

\vspace{-0.2cm}
\begin{table}[h!]
\centering
\footnotesize
\caption{Comparison of SASV systems with proposed score fusion.}
\begin{tabular}{|c|c|c|} 
\hline
SASV System & \#Trainable Params & min a-DCF \\ \hline
Embedding Fusion & 180736 & 0.1234 \\
\hline
\textbf{L-SF} & 0 & 0.5311 \\
\hline
\textbf{L-Cal-SF} & 4 & 0.0648 \\
\hline
\textbf{NL-Cal-SF}  & 4 & 0.0508 \\ \hline
\textbf{aDCF-NL-Cal-SF} & 4 & 0.0289 \\ \hline
\end{tabular}
\vspace{-0.5cm}
\label{tab:score_fusion_tab}
\end{table}

\section{Conclusion}
\vspace{-0.1cm}

In this study, we demonstrated that an SASV system either embedding or score fusion can be effectively optimized for the new a-DCF metric. Incorporating both a-DCF and BCE losses in the optimization improved performance for embedding fusion over the baseline system. Importantly, the usage of a novel threshold optimization technique further enhanced the system's capabilities, resulting in improvement in a-DCF values. These findings highlight the potential for adjusting ASV systems to achieve an optimal balance between user convenience and security against to evolving spoofing attacks.




\bibliographystyle{IEEEtran}
\bibliography{bibliography}

\end{document}